\documentclass[10pt,journal]{IEEEtran}

\usepackage{outline}
\usepackage{pmgraph}
\usepackage[normalem]{ulem}
\usepackage[utf8]{inputenc}
\usepackage{amssymb}
\usepackage{hyperref}
\usepackage{amsmath}
\usepackage{graphicx}
\usepackage{times}
\usepackage{xcolor}
\usepackage{xspace}
\usepackage[colorinlistoftodos]{todonotes} 
\usepackage{cite}
\usepackage{bm} 
\usepackage{url}

\usepackage{epstopdf}
\AppendGraphicsExtensions{.tiff}
\graphicspath{{fig/}} 

\usepackage{epsfig}
\usepackage{tikz}
\usetikzlibrary{spy}
\usepackage{algpseudocode}
\usepackage{algorithm}
\usepackage{mathrsfs}

\def\QED{~\rule[-1pt]{5pt}{5pt}\par\medskip}

\long\def\comment#1{} 

\newcommand{\xmath}[1] {\ensuremath{#1}\xspace}
\newcommand{\blmath}[1] {\xmath{\bm{#1}}}

\newcommand{\yb}{{\blmath y}}

\newcommand{\Xc}{\mathcal{X}}
\newcommand{\Yc}{\mathcal{Y}}

\newcommand{\Rd}{{\mathbb R}}

\newcommand{\Zc}{{{\mathcal Z}}}

\newcommand{\beq}{\begin{equation}}
\newcommand{\eeq}{\end{equation}}
\newcommand{\beqa}{\begin{eqnarray}}
\newcommand{\eeqa}{\end{eqnarray}}

\usepackage{multirow}
\usepackage{makecell}
\usepackage{booktabs}
\usepackage{etoolbox}
\AtBeginEnvironment{figure}{\setlength{\intextsep}{0.5\baselineskip}}

\begin{document}
\title{OT-driven Multi-Domain Unsupervised Ultrasound Image Artifact Removal using a Single CNN}
\author{Jaeyoung Huh, Shujaat Khan, and Jong Chul Ye, \IEEEmembership{Fellow, IEEE}
\thanks{The authors are with the Department of Bio and Brain Engineering, Korea Advanced Institute of Science and Technology (KAIST), 
		Daejeon 34141, Republic of Korea (e-mail:\{woori93,shujaat,jong.ye\}@kaist.ac.kr). 
		This work was supported by the National Research Foundation
of Korea under Grant NRF-2020R1A2B5B03001980.
}
}

\maketitle

\begin{abstract}
Ultrasound imaging (US)   often suffers from distinct image artifacts  from various sources.
Classic approaches for solving these problems are usually model-based iterative approaches that have been developed specifically for each type of artifact, which are often computationally intensive.
  Recently, deep learning approaches have been proposed as  computationally efficient and high performance alternatives.
Unfortunately, in the current deep learning approaches, a dedicated neural network should be trained with matched training data for each specific artifact type. 
This poses a fundamental limitation in the practical
use of deep learning for US, since large number of models should be stored to deal with various US image artifacts.
Inspired by the recent success of multi-domain image transfer, here we propose a novel, {unsupervised}, deep learning approach in which a single
neural network can be used to deal with different types of US artifacts  simply  by changing  a mask vector that switches between different target domains.
Our algorithm is rigorously derived using an optimal transport (OT) theory for cascaded probability measures. 
Experimental results using phantom and in vivo data 
demonstrate that the proposed method can generate high quality image by removing 
 distinct artifacts, which are comparable to those obtained by separately
trained multiple neural networks.
\end{abstract}


\section{Introduction}
\label{sec:introduction}
\IEEEPARstart{I}n contrast to computed tomography (CT) and magnetic resonance imaging (MRI),
ultrasound imaging (US) poses no radiation risks to the patient and enjoys fast acquisition time,  while the hardware system is much simpler. Therefore, US is very useful for many clinical and portable diagnostic applications.

Unfortunately, US suffers from various imaging artifacts such as speckle noises from signal interference, image
blur from limited  bandwidth and aperture size, etc. For the last few decades, many researchers have proposed
various model-based iterative algorithms to address these problems \cite{duan2016increasing,jirik2008two,NLLR_Despeckle,coupe2009nonlocal}.  While the results are impressive, one of the limitations of these model-based approaches is
that computationally extensive optimization problem
should be solved again for each measurement.

\begin{figure}[h!]
\center
	\includegraphics[width=9cm]{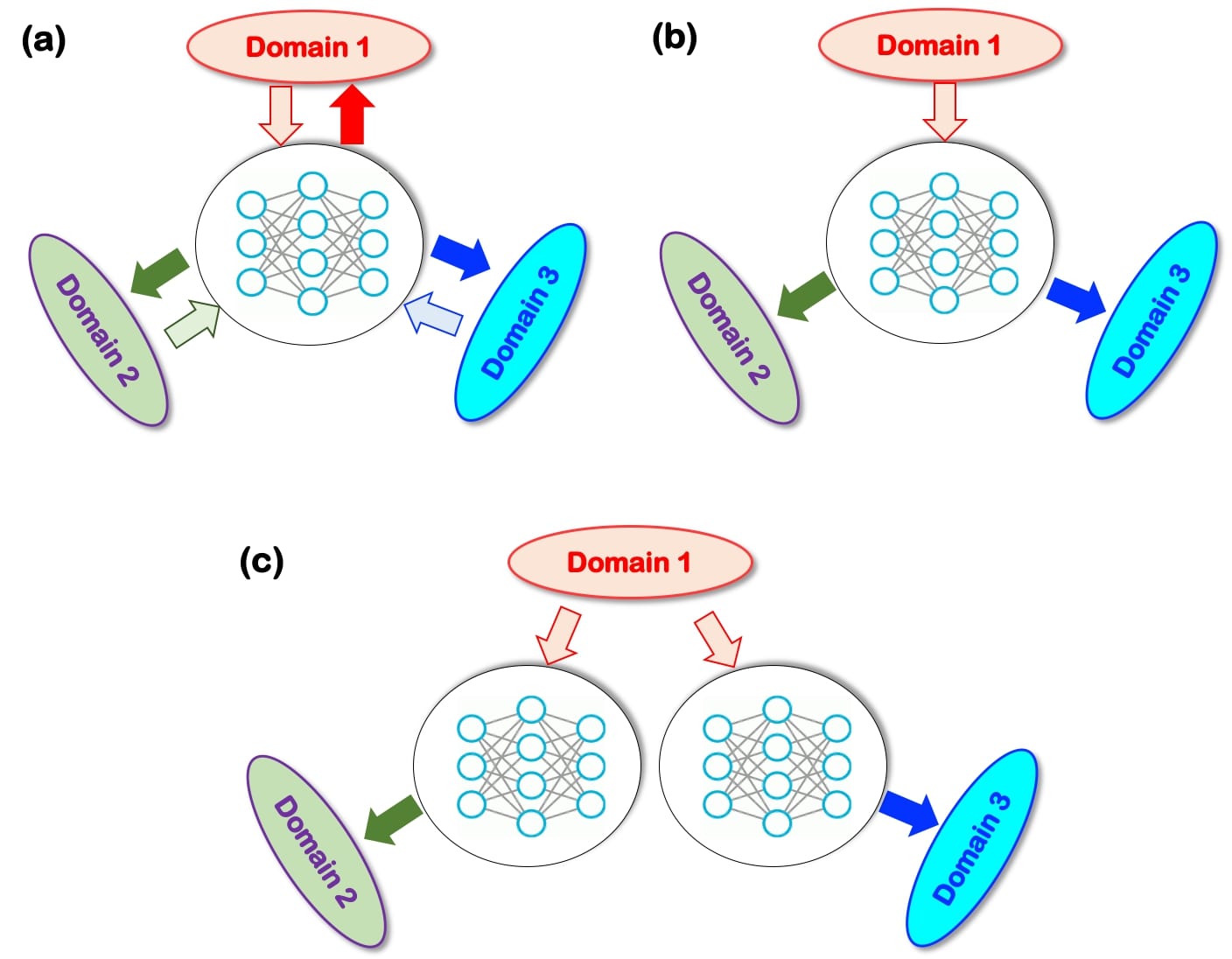}
	\vspace*{-0.5cm}
	\caption{Multi-domain image translation using (a) StarGAN,  (b) our method, and (c) CycleGAN.}
	\label{fig:concept}
\end{figure}

%
Recently, deep learning approaches has been widely used  for various medical imaging problems,
such as low-dose CT \cite{kang2017deep},
 accelerated MRI \cite{schlemper2017deep,lee2019k}, etc. 
In US,  Yoon et al \cite{yoon2019efficient} proposed a convolutional neural  network (CNN) approach to interpolate subsampled RF channel data. 
Deep learning-based beamformers have been also suggested as  promising alternatives to the delay-and-sum (DAS) or adaptive beamformers \cite{khan2020adaptive}. Furthermore, various US artifact removal
algorithms have been  implemented using deep neural networks \cite{khan2020adaptive,kokil2020despeckling}.
These deep learning approaches are  {\em inductive} in the sense that  the trained network can be used for other measurement data  without 
additional optimization procedure. Therefore, deep learning approaches are ideally suitable for medical imaging problems,
where fast reconstruction is critical.

In spite of these pioneering works,  the deep neural network approaches for US imaging still have several
technical huddles for their wide acceptance.  First, US images are usually corrupted with different types of artifacts, and each user often prefers
distinct choice of artifact suppression algorithms depending on the clinical applications.
This implies that large number of neural networks should be stored in a  US scanner to deal with various image artifacts
to satisfy the users' demands.
Second, most of the current deep learning approaches are based
on supervised training. Although this requires matched ground-truth artifact-free images,
 obtaining matched artifact-free image  is challenging
in ultrasound imaging.

In this paper, we propose a novel deep learning approach to overcome these fundamental
limitations.
In our method, a  {\em single} neural network can be switched  to address different types of image artifacts by simply
changing the target mask vector.
Moreover, our neural network is trained
in an unsupervised manner so that it does not require any matched reference data for training. 

In fact, this is inspired by the recent successes of
multi-domain image transfer in computer vision literature \cite{choi2018stargan,lee2019collagan}.
For example, 
Cho et al \cite{choi2018stargan}  proposed a StarGAN, which uses only one generator with one discriminator  to translate an image
to multiple domains by
utilizing a mask vector to indicate which domain to transfer.
However, as will be shown later, we found that the direct
application of StarGAN is not necessary,  since the US  image artifacts removal problem  is a uni-directional
translation problem.  More specifically, while 
StarGAN should translate  any domain to another one equally well as shown in Fig.~\ref{fig:concept}(a), 
this is not necessary in US artifact
removal problems, since  DAS images are raw data obtained from the scanner and
it is not necessary to re-generate DAS images from other domain  images.
Therefore,  a correct image reconstruction pipeline should be asymmetric as shown in 
Fig.~\ref{fig:concept}(b), where artifact suppressed  images (domain 2 and 3) are generated only from DAS raw data (domain 1).
It turns out that the lack of symmetry makes the multi-domain translation task much simpler,
thereby
significantly reducing the network  complexity and improving the performance.

Another important contribution of this work is a novel geometric insight that leads to the
proposed network architecture.  Specifically, inspired by the recent theory of optimal
transport driven cycleGAN \cite{sim2019optimal}, we demonstrate that
the multi-domain artifact removal problem can be formulated as an optimal transport problems
between cascaded probability measures, where the goal is to minimize the statistical distances between the
``push-forward'' measures and the real data distributions. It turns out that if Wasserstein-1
metric\cite{villani2008optimal} and Kullback-Leiber (KL) divergence are used as distances for continuous and discrete
distributions, respectively, then the proposed multi-domain unsupervised image translation network emerges.


 \section{Backgrounds}
\label{sec:review}


\subsection{Multi-Domain Image Translation}


One of the earliest works in image translation is Pix2Pix \cite{isola2017image} based on conditional generative adversarial networks (cGAN). 
Pix2pix was designed for various translation tasks with paired data such as  map to aerial photo, or sketch to photo, and so on \cite{isola2017image}.

To relieve the stringent requirement of paired data, Zhu et al \cite{zhu2017unpaired} proposed an unsupervised image-translation
technique called CycleGAN, which consists of two generators and two discriminators.
Specifically, one generator is used to generate a target domain image and the other generator is used to generate the original domain image.
Then, the cycle-consistency loss is used to impose the consistency of the 
 the input image and the regenerated fake image from the second generator. 
 CycleGAN has been successfully used as a promising unsupervised learning approach for various medical imaging problems,
  such as low-dose CT \cite{kang2019cycle},   etc.


However, CycleGAN approach is not scalable in the sense  that a total of $ N (N-1) $ CNN generators are required to translate between $ N $ domains  \cite {choi2018stargan}.
To deal with the scalability issue of CycleGAN in multi-domain image translation,
Choi et al \cite {choi2018stargan} proposed a StarGAN approach which performs multi-domain image translation with only one generator and one discriminator. The domain is separated with a mask vector, which consists of one-hot vector. 
The target domain mask vector then guides the translation between different image domains.
Moreover, the discriminator is designed not only to differentiate between real and fake, but also to classify which domain the image belongs to.

By extending this idea furthermore, Collaborative GAN (CollaGAN) \cite {lee2019collagan} 
was proposed to deal with the multi-domain transfer from multiple inputs.
In particular, CollaGAN
uses multiple inputs to generate
more realistic target domain images, by exploiting the multiple cycle-consistency loss. 
Similar to StarGAN, CollaGAN is also based on single generator and discriminator.

One of the main assumptions in the aforementioned image translation tasks is that each domain is equally important, i.e. an image in any domain should be translated to another domain equally well regardless of source and target domain combination.
In fact, this turns out to be quite stringent requirement in the network training.
As will be discussed later, in US artifact removal problems the raw data always belongs to one specific domain and the goal is to process
the raw data using different types of algorithms. Therefore, the translation task is uni-directional, which could provide an opportunity to relax
the stringent constraint for network training.

\subsection{Optimal Transport driven CycleGAN}

As the multi-domain image translation is based on the generative adversarial network (GAN)\cite{choi2018stargan,lee2019collagan}, one may wonder whether
the image translation is just a cosmetic change or real.
In fact, this is a important issue in medical imaging applications, 
since any artificially created features
can reduce the diagnostic accuracy of the medical images.


%
\begin{figure}[!hbt] 	
\center{ 
\includegraphics[width=7cm]{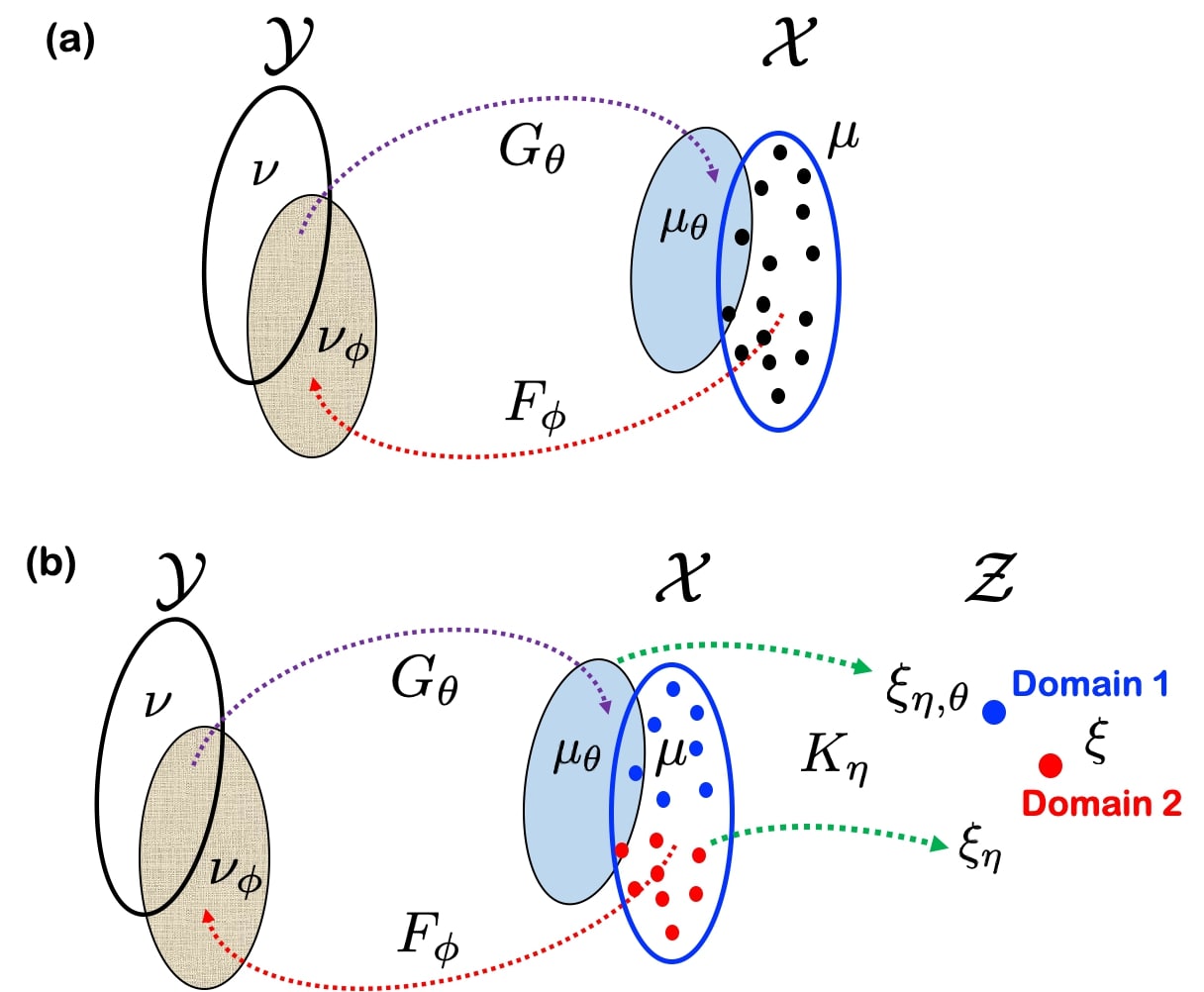}
}
\caption{Geometric view  of (a) CycleGAN, and (b) the proposed multi-domain transfer.}
\label{fig:geometry}
\end{figure}

In our recent  paper \cite{sim2019optimal}, we showed that the CycleGAN architecture  can be rigorously derived
from an optimal transport problem \cite{villani2008optimal} of one probability distribution to another.
Accordingly, we can safely assume that the generated image reflects real image distribution if networks are properly
designed and trained.
As our goal is to extend this to multi-domain translation, in the following we briefly review OT driven CycleGAN \cite{sim2019optimal} for self-containment.

{Optimal transport (OT) provides a mathematical means to operate between two probability measures} \cite{villani2008optimal}. 
 Formally, we say that  $T:\Xc \mapsto \Yc$ transports a probability measure (or distribution)  $\mu \in P(\Xc)$ to another measure $\nu \in  P(\Yc)$, if
\begin{eqnarray}\label{eq:constraint}
\nu(B) = \mu\left(T^{-1}(B)\right),\quad \mbox{for all $\nu$-measurable sets $B$},
\end{eqnarray}
which is often denoted as the {\em push-forward} of the measure, and represented by
$$\nu = T_\#\mu .$$ 
{Suppose there is a cost function $c:\Xc \times \Yc \rightarrow \Rd\cup\{\infty\}$ such that $c(x,y)$ represents the cost of moving one unit of mass from $x \in \Xc$ to $y \in \Yc$.}
The OT problem \cite{villani2008optimal} is then to find a transport map $T$ that transports $\mu$ to $\nu$
at the minimum total transportation cost.

Now, we will explain how optimal transport can lead to a CycleGAN architecture \cite{sim2019optimal}.
Suppose that the target image space
$\Xc$ is equipped with a probability measure $\mu$, whereas
the original image space is $\Yc$ with a probability measure $\nu$.
In unsupervised learning problem,  there are no paired data, so
 the goal of unsupervised learning is to match the probability distributions rather than each individual sample.
This can be done by finding  transportation maps that transport the measure $\mu$ to $\nu$, and vice versa.

More specifically, as shown in Fig.~\ref{fig:geometry}(a),
 the transportation from a measure space $(\Yc,\nu)$ to another measure space $(\Xc,\mu)$ is done by a generator $G_\theta: \Yc \mapsto \Xc$, realized by a deep
network parameterized with $\theta$.  Then, the generator $G_\theta$ pushes forward the measure $\nu$ in $\Yc$ to a measure $\mu_\theta$ in the target space $\Xc$, i.e. $\mu_\theta=G_{\theta\#}\nu$. Similarly, the transport from $(\Xc,\mu)$ to $(\Yc,\nu)$ is performed by another neural network generator $F_\phi$,
 so that the generator $F_\phi$ pushes forward the measure $\mu$ in $\Xc$ to $\nu_\phi$ in the original space $\Yc$.
Then, our goal is to derive an optimal transport map by minimizing the statistical distances between $\mu$ and $\mu_\theta$, and
 between $\nu$ and $\nu_\phi$, respectively.
 
One of the most important contributions of our work \cite{sim2019optimal} is to show  that if the statistical distance is measured by Wasserstein-1 metric \cite{villani2008optimal} and the distance minimization
 problem is solved using a dual formulation, a CycleGAN architecture emerges \cite{sim2019optimal}.
 This gives an important mathematical foundation of CycleGAN.

\section{Problem Formulation}
\label{sec:problem}


\subsection{DAS Image Artifacts}

In US imaging, 
the preprocessed channel data is first obtained from the return RF echoes:
 \begin{eqnarray}
 \yb_{l,n}=\begin{bmatrix} y_{l,n}[0] & y_{l,n}[1] & \cdots &y_{l,n}[{N}-1]\end{bmatrix}^\top \in \Rd^{N}
 \end{eqnarray}
where $y_{l,n}[c]$ denotes the delay corrected RF data at the $c$-th channel element, and
 $l$ and $n$ denote the scan line index and the depth, and $N$ is the aperture size.
The   delay and sum (DAS) beamformer for the $l$-th scanline at the depth sample $n$ can be then computed as 
\begin{equation}\label{eq:DAS}
{x}_{l,n} 
=\frac{1}{N}\mathbf{1}^\top\yb_{l,n} 
\end{equation}
where $\mathbf{1}$ denotes a $N$-dimensional column-vector of ones.

The DAS beamformer is designed to extract the  low-frequency spatial content that corresponds to the energy within the main lobe; thus,
large number of receiver elements are often necessary in DAS beamformer to improve the image quality by reducing the side lobes. Moreover, to calculate accurate time delay, sufficiently large bandwidth transducers are required.
Therefore, in many US systems, especially for low-end or portable systems where
the number of receiver elements is not sufficient and transducer bandwidth is limited,
the image quality degradation is unavoidable.

Another  performance degradation of 
DAS beamformer comes from the inaccuracy of ray approximation of the wave propagation,
as the underlying image physics is from the wave phenomenon and the deviation from the
real physical introduces inevitable image artifacts.

Yet another image artifacts often observed in DAS image is speckle noise. Speckles noises are originated
from the small size scatterers which reflects the US energy. As the DAS simply adds the return echoes after proper time delay,
the constructive and destructive interferences from these scattered echoes generate bright and dark spots in the image, which results in the speckle
noise. For example,  for cardiac volume quantification
using segmentation, speckle noise imposes significant algorithmic challenges, so the proper suppression of the speckle noise is often
necessary.
That said, in some clinical applications such as liver imaging, the speckle statistics is an important
measure for the diagnosis of diseases progress, because it is highly sensitive to small variations in scatterers. 
Accordingly, the speckle noise removal algorithm should be used judiciously depending on specific applications.

\subsection{Classical Image Processing Pipelines}

In the following, we will describe classical model-based iterative methods in more detail. 

\subsubsection{Deconvolution}
\label{sec:classical_deconv}


In US, the received signal is modeled as a convolution of tissue reflectivity function (TRF) $x$ with a point spread function (PSF) $h$, where tissue reflectivity function represents scatter's acoustic properties, while the impulse response of the imaging system is modeled by point spread function. 
In most practical cases, the complete knowledge of $h$ is not available, and therefore both unknown TRF $x$ and  the PSF $h$ have to be estimated together. These types of problems are often called deconvolution problems.
Deconvolution ultrasound may help in dealing with modeling inaccuracies and finite bandwidth issues, which will eventually improve the spatial resolution of an ultrasonic imaging system \cite{duan2016increasing,jirik2008two}.  

Mathematically, the  deconvolution problem can be formulated as follows:
\begin{align}\label{eq:deconv}
\hat{x} =\arg\min_{x,h} ||y-h\ast x||^{2}+\lambda R(x) 
\end{align}
where $R(x)$ is a sparsity imposing regularization terms such as $l_1$, total variation (TV), etc.
One strategy to address \eqref{eq:deconv} is to estimate $h$ and $x$ jointly, and another strategy is to estimate them separately, i.e., first $h$ is estimated from $y$, and then $x$ is estimated based on $h$ \cite{jirik2008two}. 
In this paper,  the second strategy is used for the generation of unmatched target data distribution,
i.e. images in Domain 2 in Fig.~\ref{fig:concept}(b).

\subsubsection{Despeckle}
\label{sec:classical_despeckle}

Speckle noises  are one of the major reasons of quality degradation, so that ``despeckle'' (removal of the speckles) can improve the visual quality and subsequently enhance the structural details in US images \cite{NLLR_Despeckle}. In recent past, a variety of reasonably good de-speckling algorithms have been proposed for US imaging \cite{NLLR_Despeckle, coupe2009nonlocal}.  One such algorithm is proposed by Zhu \textit{et al} \cite{NLLR_Despeckle}, which is based on the principle of non-local low-rank (NLLR) filtering.

In NLLR \cite{NLLR_Despeckle},  the guidance image is first obtained as a reference.
Then, using guidance images, group of similar patches are selected based on the similarity to the reference patch. 
Using the group of similar patches, the low-rank + sparse decomposition \cite{candes2011robust} is then performed.
The optimization problem is solved using alternating directional method of multiplier (ADMM) \cite{yin2008bregman}. 
Finally, the low-rank component is considered as speckle free images, whereas the sparse component are considered as speckle noises.

Although the algorithm provides an effective
means for removing speckle, it is computationally expensive, so it cannot be used for routine US applications.
In this paper, this algorithm is used to generate our target samples for despeckle images (i.e., images in Domain 3 in Fig.~\ref{fig:concept}(b)).

\subsection {Multi-domain Unsupervised Artifact Removal}

In this paper,  we are interested in  a single neural network that can remove both blur artifacts and speckle noises
from the DAS images in a real-time manner.  
More specifically,  in  Fig.~\ref{fig:concept}(b), Domain 1 is for the DAS input images,
and Domain 2 and 3 correspond to deconvolution and despeckle image domains.
Since the use of deconvolution images or despeckle images depends on specific clinical
applications, our goal is therefore  to design a neural network such that  it
can be switched to process different type of artifacts by mere changing the target domain mask vector.

However, if the existing CycleGAN is used for this applications,
 two separate
neural networks are required for each artifact type as shown in Fig.~\ref{fig:concept}(c), which increases additional
burden for US scanner.
%

\begin{figure}[h]
\center
	\includegraphics[width=0.5\textwidth]{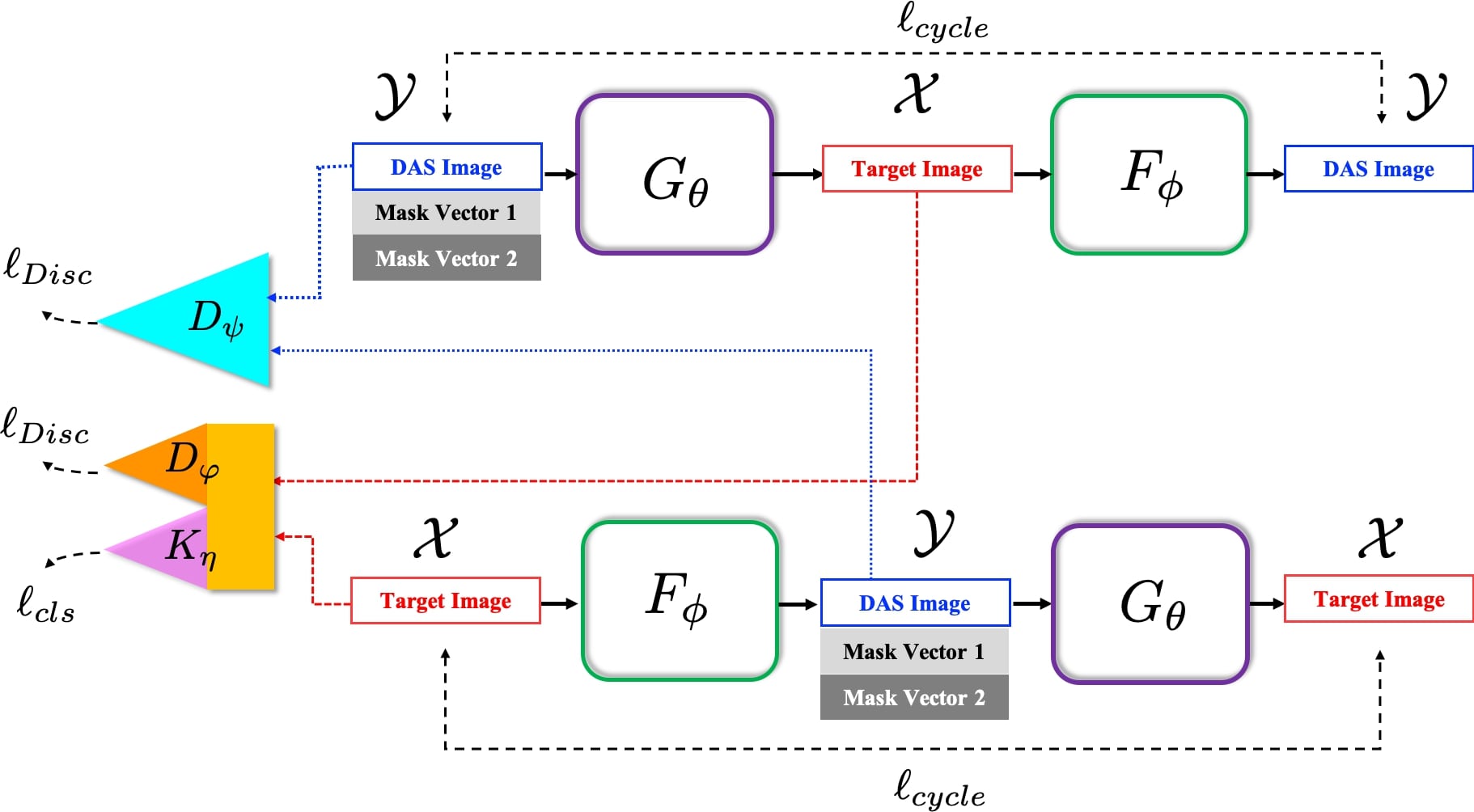}
	\vspace*{-0.5cm}
	\caption{Proposed multi-domain unsupervised artifact removal networks.}
	\label{fig:scheme}
\end{figure}

\section{Main Contribution}
\label{sec:theory}


\subsection{Geometric Formulation}

The geometry of our  problem is illustrated in
Fig.~\ref{fig:geometry}(b), which is similar to that of CycleGAN in 
Fig.~\ref{fig:geometry}(a) but with important differences.

Here, the original image space $\Yc$ is the DAS image domain, whose probability distribution  is $\nu$,
whereas the target image space
$\Xc$ contains both despeckle  images and deconvolution images, and their distribution
follows the probability measure $\mu$.
One key difference compared to the CycleGAN geometry in Fig.~\ref{fig:geometry}(a) is that
the target distribution $\mu$ is modeled as the mixture distribution of despeckle  and deconvolution images,
which can be separately by a classifier $K_\eta: \Xc\mapsto \Zc$ that maps each sample
in $\Xc$ to discrete domain labels in $\Zc$ with a discrete measure $\xi$.

Thus, our goal is to find  a generator $G_\theta: \Yc \mapsto \Xc$ that
transport $\nu$ to a probability measure $\mu_\theta$ which is close to $\mu$ and still enables the correct
classification using $K_\eta$. 
Accordingly, $G_\theta$ and $K_\eta$ are dependent to each other, so they should be optimized simultaneously.
In fact, this corresponds to an optimal transport problem to a cascaded measures.
On the other hand,  the generator $F_\phi$ that transport $\mu$ to $\nu_\theta$  should transport
all mixture distribution to the probability distribution $\nu$ in the original space $\Yc$.

%
 
%

Similar to the optimal transport driven CycleGAN \cite{sim2019optimal}, in our method the statistical distance is measured
using the Wasserstein-1 metric.
More specifically,
for the choice of a metric $d(x,x')=\|x-x'\|$ in $\Xc$,  
the Wasserstein-1 between $\mu$ and $\mu_\theta$ can
be computed by:
\begin{align}\label{eq:Wmu}
W_1(\mu,\mu_\theta)
=&\inf\limits_{\pi \in \Pi(\mu,\nu)}\int_{\Xc\times \Yc} \|x-G_\theta(y)\|d\pi(x,y) 
\end{align}
where $\Pi(\mu,\nu)$ denotes the set of joint distribution whose marginal distribution is $\mu$ and $\nu$.
Similarly, the Wasserstein-1 distance between $\nu$ and $\nu_\phi$ is given by
\begin{align}\label{eq:Wnu}
W_1(\nu,\nu_\phi)
=&\inf\limits_{\pi \in \Pi(\mu,\nu)}\int_{\Xc\times \Yc} \|F_\phi(x)-y\|d\pi(x,y) 
\end{align}
Since the optimal joint distribution $\pi$ minimizing \eqref{eq:Wmu} and \eqref{eq:Wnu}
could be different if they are minimized separately, 
a better way of finding the transportation map is to minimize them together with the same joint distribution $\pi$:
\begin{align}\label{eq:unsupervised}
\inf\limits_{\pi \in \Pi(\mu,\nu)}\int_{\Xc\times \Yc} \|x-G_\theta(y)\|+ \|F_\phi(x)-y\|d\pi(x,y) 
\end{align}
This can be considered as the sum of the statistical distances in both spaces.

Next, to enable a multi-domain transfer,  the measure in the target space $\Xc$ should be
transported to the discrete sample space $\Zc$ with the discrete measure $\xi$.
As we have two distribution $\mu$ and $\mu_\theta$ in $\Xc$, the corresponding push forward
measures by the classifier $K_\eta$ are given by $\xi_\eta=K_{\eta\#}\mu$ and
$\xi_{\eta,\theta}=K_{\eta\#}\mu_\theta$, respectivey (see Fig.~\ref{fig:geometry}(b)).
Then, the classification design problem can be  done again by minimizing the statistical distances in $\Zc$. 

Since
the classification label is discrete (in our case, binary),  one of the simplest way of
measuring the statistical distance in $\Zc$ is using the Kullback-Leiber (KL) divergence.
Specifically, the KL divergence between
$\xi$ and the push-forward measure $\xi_\eta$ is given by:
\begin{align*}
D_{KL}(\xi||\xi_\eta)  
 &= \int \log \left(\frac{p(x)}{p_\eta(x)}\right) p(x) dx \\
  &= -\int p(x) \log {K_\eta(x)}  dx + \mbox{const.}
 \end{align*}
 where  the output of the classifier $K_\eta$ is the probability
 distribution indicating the probability belonging to  target domains,  
 and $p(x)$ is the one-hot vector encoded true label probability for $x$.
 For example, in the conversion of DAS image to two target domains (decovolution, and despeckle),
 $p(x)$ is a discrete probability mass function which is $p_1(x)=1, p_2(x)=0$ for the
 deconvolution target, and $p_1(x)=0, p_2(x)=1$ for despeckle targets.

 Similarly, the KL divergence between
$\xi$ and the push-forward measure $\xi_{\eta,\theta}$ is given by:
\begin{align}\label{eq:KL2}
D_{KL}(\xi||\xi_{\eta,\theta})  
  &= -\int p(G_\theta(y)) \log K_\eta(G_\theta(y)) dy + \mbox{const.}
 \end{align}
 where $p(G_\theta(y))$  is one-hot vector encoded target label for the generator $G_\theta$ for
 a given DAS image $y$.

By putting them together, the overall loss becomes
\begin{align}
&\ell_{total}(\theta,\phi,\eta)  \label{eq:loss} \\
= &\inf\limits_{\pi \in \Pi(\mu,\nu)}\int_{\Xc\times \Yc} \|x-G_\theta(y)\|+ \|F_\phi(x)-y\|d\pi(x,y)  \notag \\
-&\lambda_{cls}\left(\int_\Xc p(x) \log {K_\eta(x)}  dx\right. \notag \\
& \left.+\int_\Yc p(G_\theta(y)) \log {K_\eta(G_\theta(y))}  dy \right)\notag
\end{align}
where $\lambda_{cls}$ is the weighting parameter for the classifier.

\subsection{Proposed Method}

Unfortunately, direct minimization of the loss in \eqref{eq:loss} is difficult specially due to the need for non-parametric
minimization with respect to the joint distribution $\pi$.
To address this, one of the most important contributions of our companion paper \cite{sim2019optimal} is to show that
the following primal problem
\begin{align}\label{eq:primal}
\inf\limits_{\pi \in \Pi(\mu,\nu)}\int_{\Xc\times \Yc} \|x-G_\theta(y)\|+ \|F_\phi(x)-y\|d\pi(x,y) 
\end{align}
is equivalent to the following dual formulation:
\begin{eqnarray}\label{eq:OTcycleGAN}
\max_{\psi,\varphi}\ell_{cycleGAN}(\theta,\phi;\psi,\varphi)
\end{eqnarray}
where 
\begin{eqnarray}
\ell_{cycleGAN}(\theta,\phi):=  \lambda_{cyc} \ell_{cycle}(\theta,\phi) +\ell_{Disc}(\theta,\phi;\psi,\varphi) 
\end{eqnarray}
where $\lambda_{cyc}>0$ is the hyper-parameter, and  the cycle-consistency term is given by
\begin{align*}
\ell_{cycle}(\theta,\phi)  =& \int_{\Xc} \|x- G_\theta(F_\phi(x)) \|  d\mu(x) \\
&+\int_{\Yc} \|y-F_\phi(G_\theta(y))\|   d\nu(y)
\end{align*}
whereas  the discriminator term is
\begin{align*}
&\ell_{Disc}(\theta,\phi;\psi,\varphi)  \\
=&\max_{\varphi}\int_\Xc D_\varphi(x)  d\mu(x) - \int_\Yc D_\varphi(G_\theta(y))d\nu(y)  \\
 & + \max_{\psi}\int_{\Yc} D_\psi(y)  d\nu(y) - \int_\Xc D_\psi(F_\phi(x))  d\mu(x) \notag
\end{align*}
Here, $D_\varphi,D_\psi$  should satisfy 1-Lipschitz condition (i.e.
\begin{align*}
|D_\varphi(x)-D_\varphi(x')|\leq \|x-x'\|,&~\forall x,x'\in \Xc \\
|D_\psi(y)-D_\psi(y')|\leq \|y-y'\|,&~\forall y,y'\in \Yc
\end{align*}
To impose 1-Lipschitz condition, 
the gradient of the Kantorovich potential is constrained to be 1 \cite{gulrajani2017improved}:
\begin{align*}
&\ell_{GP}(\varphi,\psi)= \\
&-\int_{{\Xc}}(\|\nabla_{\tilde x}D_\varphi(x)\|_2 - 1)^2d\mu(x) -\int_{{\Yc}}(\|\nabla_{\tilde y}D_\psi(y)\|_2 - 1)^2d\nu(y)  \notag
\end{align*}
where 
  $\tilde x=\alpha x+(1 - \alpha)G_\theta(y)$ and   $\tilde y=\alpha y+(1 - \alpha)F_\phi(x)$ with $\alpha$ 
being random variables from the uniform distribution between $[0,1]$ \cite{gulrajani2017improved}.

By collecting all terms together, our multi-domain artifact removal problem can be formulated by
\begin{align}\label{eq:final}
\min_{\theta,\phi,\eta}\max_{\varphi,\psi} \ell_{mlt}(\theta,\phi,\eta;\varphi,\psi)
\end{align}
where
\begin{align*}
&\ell_{mlt}(\theta,\phi,\eta;\varphi,\psi)\\
:= & 
 \lambda_{cyc} \ell_{cycle}(\theta,\phi) +\ell_{Disc}(\theta,\phi;\psi,\varphi)  + \lambda_{GP} \ell_{GP}(\varphi,\psi)  \\
-&\lambda_{cls}\left(\int_\Xc p(x) \log {K_\eta(x)}  dx\right. \notag \\
& \left.+\int_\Yc p(G_\theta(y)) \log {K_\eta(G_\theta(y))}  dy \right)\notag
\end{align*}
where $\lambda_{GP}$, $\lambda_{cls}$, $\lambda_{cyc}$ are the  weighting parameters depending on the relative importance. 
In this paper, we used $\lambda_{GP}=30$, $\lambda_{cls}=1$, and $\lambda_{cyc}=20$.

\subsection {Neural Network Implementation}

Fig. \ref{fig:scheme} illustrates the proposed multi-domain unsupervised artifact removal
network originated from the optimization problem in \eqref{eq:final}.  The network architecture consists of two generators. 
 The main generator $G_\theta$ 
 converts DAS images to either deconvolution or de-speckled images.
The second generator $F_\phi$ returns the  deconvolution or despeckle images  to the original DAS domain images. 
Once the networks are trained, note that we only use $G_\theta$ at the inference phase.

Here, care should be taken due to the existence of the classifier.
More specifically, in calculating the KL divergence in \eqref{eq:KL2}, 
the target domain of $G_\theta(y)$ should be indicated.
Therefore, the main idea is to indicate the target domain using an input mask vector, $m$:
  \begin{align}
x = G_\theta(y; m)
  \end{align}
  For example, a  DAS input image with the mask vector $m=[1,0]$ creates the deconvolution image, whereas
   an input DAS image with the mask vector $m=[0,1]$  is converted to a despeckled image. The mask vector is augmented along the channel direction  with the input image. 
  On the other hand, $F_\phi$ does not need a mask vector since there exists no classifier in $\Yc$.
  
  \begin{figure}[h!]
  \center
	\includegraphics[width=8cm]{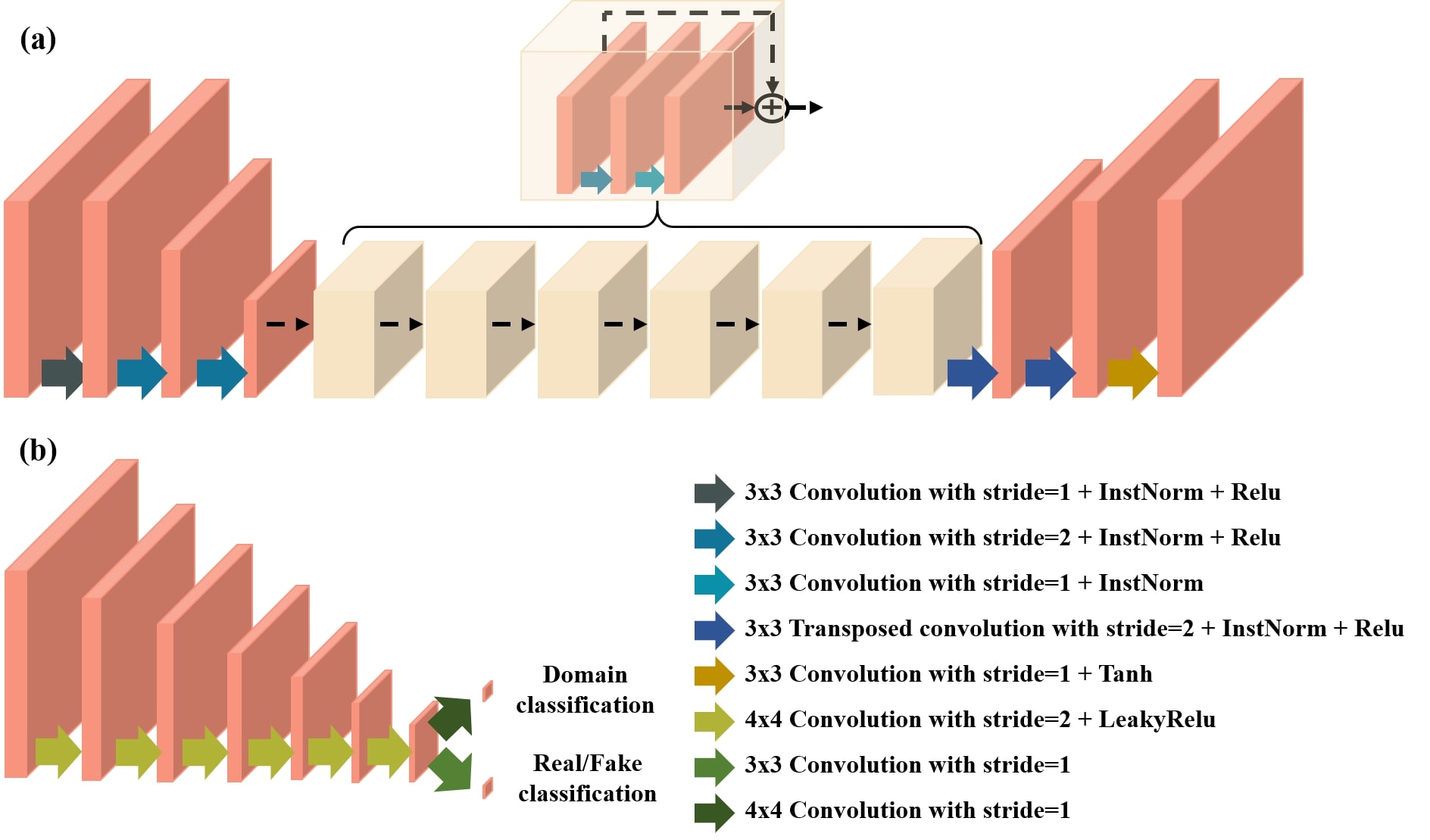}\
	\caption{Network architecture: (a) generator, and  (b) discriminator/domain classifier.}
	\label{fig:network_architecture}
\end{figure}

Accordingly, although the same  generator architecture as shown in Fig.~\ref{fig:network_architecture}(a)
is used for both $G_\theta$ and $F_\phi$,  the input channel numbers are different.
For the case of $G_\theta$, the number of input channels is 3, which is composed of DAS images and two mask vector channels,
whereas $F_\phi$ is composed of a single input channel.
  
Additionally, there are also  two discriminators $D_\varphi$ and $D_\psi$ as shown in Fig.~\ref{fig:scheme}. 
Specifically, 
$D_\varphi$ tries to find the difference between the true image $x$ and the generated image $G_\Theta(y)$,
whereas $D_\psi$ attempts to find the fake measurement data  that are generated by the synthetic
measurement procedure $F_\phi(x)$.

Finally, we have the domain classifier $K_\eta$ to distinguish between deconvolution and despeckled
images. In fact, the domain classifier and discriminators are both classifiers, so
their structure share many  commonalities.
Therefore, as shown in  Fig.~\ref{fig:network_architecture}(b), the discriminator
$D_\varphi$ and the classifier $K_\eta$ are implemented using a same network architecture with  double output
headers composed of a domain classifier or discriminator.

\section{Methods}
\label{sec:methods}

\subsection {Dataset}
 We used in vivo and phantom images to train the network. The in vivo images are focused B-mode images using a linear probe (L3-12H) from the US system E-CUBE 12R (Alphinion, Korea). The images were measured from 10 volunteers at four parts of the carotid and thyroid areas. It has 10 time frames for each body part. The phantom images are tissue mimicking phantoms that are also acquired with focused B-mode imaging using a linear probe. Both images are taken with a center frequency of 8.5 MHz. Each image is captured with 64 channels and 96 scan lines. The details of the probe are listed in the Table \ref{table:data}. We selected 8 volunteer data, and 304 images in in vivo images and 125 images in phantom images are used for the training data set. 
 
 To create the deconvolution images in Domain 2, we used the joint sparse deconvolution model that is iterative method along the axial direction with lateral direction constraint (Sec. \ref{sec:classical_deconv}). We also used the Non-Local-Low-Rank (NLLR) algorithm to generate the despeckle images (Sec. \ref{sec:classical_despeckle}) for Domain 3 data. We used 429 deconvolution images and 478 despeckle images to train the network as target domain images. In addition, we randomly mixed the input dataset and the target dataset for each epoch separately in order to train the network with unpaired method.
 \begin{table}[h!]
	\centering
	\vspace*{-0.5cm}
		\caption{Data configuration}
			\resizebox{0.25\textwidth}{!}{
	\begin{tabular}{c|c}
		\hline
		Parameter & Configuration \\
		\hline \hline
	  	Probe Model & L3-12H\\
	  	Center frequency & 8.5MHz \\
	  	Sampling frequency & 40MHz\\
	  	Wave mode & Focused \\
	  	Element & 192 \\
	  	Tx element & 128 \\
	  	Transmision emission & 96 \\
	  	Rx elements & 64\\
	  	Element pitch & 0.2mm \\
	  	Element width & 0.14 mm \\
		\hline
	\end{tabular}
	}
		\vspace*{-0.5cm}
	\label{table:data}
\end{table}

\subsection {Network Architecture}
   As shown in Fig. \ref{fig:network_architecture}, we employed the generator architecture that was used in \cite{zhu2017unpaired,choi2018stargan}. It is composed of two downsampling steps and six residual blocks and two upsampling steps. The convolution with stride 2 is utilized for downsampling operation. The residual block has two convolution steps. The transposed convolution layer with stride 2 is used for upsampling operation. In the generator, we used instance normalization instead of batch normalization. Hyperbolic tangent (tanh) non-linearity is applied in the last layer for stablizing the training process.
   
     The discriminator is shown in Fig. \ref{fig:network_architecture} (b). We used PatchGAN which classifies the local image whether real or fake. It is simply composed of six convolution sets of 4x4 kernel size with stride 2 and LeakyReLU except last layer. In the last layer, there is multiheader structure. One head is for domain classification and the other is for real/fake classification. 

  \begin{figure*}[h]
    \center
  	\includegraphics[width=0.7\textwidth]{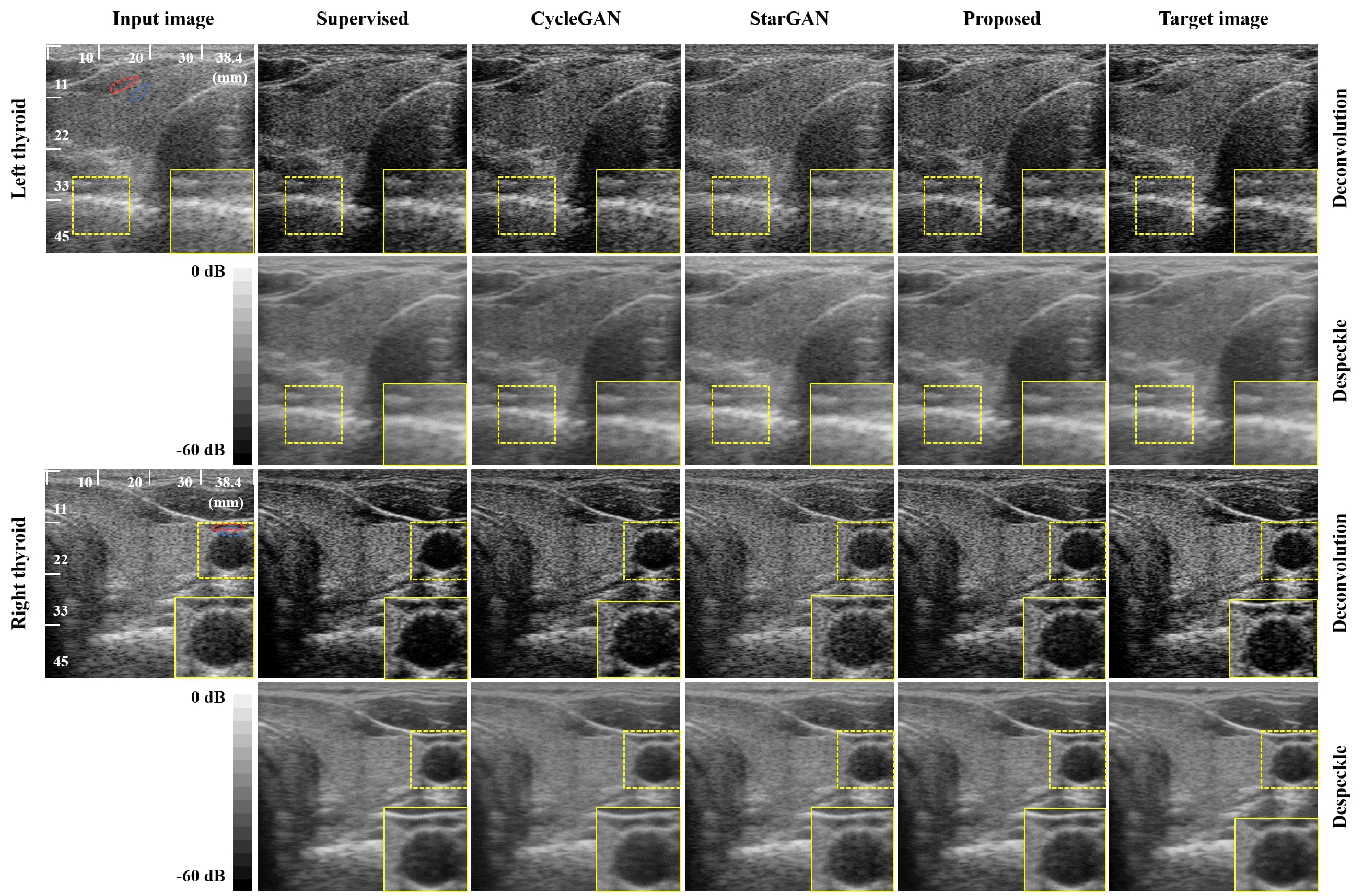}
	\vspace*{-0.3cm}
  	\caption{Comparison results with other algorithms.  The yellow box under each image shows the magnified image of yellow dot-line box. The red and blue line are the selected region for calculating contrast metrics. The first sample is left thyroid image and the second sample is the right thyroid image.}
  	\label{fig:algorithm_comparison}
  \end{figure*}
  
\subsection {Training Details}
\subsubsection {Implementation details }
To train the network, we used the Adam optimizer with $\beta=0.5$. The learning rate is started from 1e-4 and decreased linearly from the 500 epochs. The total epoch is 1000, and the batch size is 4. To avoid the overfitting problem, we used data augmentation technique like flipping, rotating and random scaling. 	We implemented all work on the Python with Tensorflow and MATLAB 2017a. We trained the network with the NVIDIA GeForce GTX 1080 Ti GPU.

\subsubsection{Evaluation metric}
To evaluate our results, we used values for the peak-signal-to-noise ratio (PSNR), the structure similarity (SSIM) \cite{1284395}, the contrast-to-noise ratio (CNR), the generalized CNR (GCNR) \cite{rodriguez2019generalized},  and the Contrast Ratio (CR). 

The PSNR value is calculated between the label image ($f$) and output image ($f'$). The equation between images whose size is $m\times n$ is
\begin{equation}
\begin{aligned}
PSNR(f, f') =  20 \log_{10}\left(\frac{nm * v_{max}}{||f-f'||_{2}}\right),
\end{aligned}
\end{equation} 
where $v_{max}$ denotes the maximum value of $f$ and $|| \cdot ||_{2}$ is $l_{2}$-norm.
The SSIM value is defined by \cite{1284395}
\begin{equation}
\begin{aligned}
SSIM(f,f') = \frac{(2\mu_{f} \mu_{f'} + c_{1})(2\sigma_{f,f'}+c_{2})}{(\mu_{f}^{2}+\mu_{f'}^{2}+c_{1})(\sigma_{f}^{2}+\sigma_{f'}^{2}+c_{2})}.
\end{aligned}
\end{equation}
The $\mu_{f}, \mu_{f'}, \sigma_{f}, \sigma_{f'}, \sigma_{f,f'}$ denotes the mean value of $f$ and $f'$, deviation value of $f$ and $f'$, covariance of $f, f'$, respectively. The $c_{1}$ is $\kappa_{1}v_{max}$ and $c_{2}$ is $ \kappa_{2}v_{max}$. In this paper, $\kappa_{1}$ and $\kappa_{2}$ are 0.01 and 0.03, respectively.
The CNR value can quantify the image contrast which is defined as
\begin{equation}
\begin{aligned}
CNR(f) 	= \frac{|\mu_{S}-\mu_{B}|}{\sqrt{\sigma_{S}^{2}+\sigma_{B}^{2}}},
\end{aligned}
\end{equation}
where 
$\mu_{S}, \mu_{B},\sigma_{S}, \sigma_{B}$ denotes the mean and standard deviation value of the structure and the background, respectively.
The GCNR value is the generalization value of CNR \cite{rodriguez2019generalized}. It calculates the area of region between intensity distribution of two areas:
\begin{equation}
\begin{aligned}
GCNR(f) = 1 - \int_{-\infty}^{\infty} \min_{x} (p_{S}(x),p_{B}(x))
\end{aligned}
\end{equation}
where $p_S$ and $p_B$ are probability distributions in structure and background, respectively.
The CR value is used to compare the contrast without noise condition:
\begin{equation}
\begin{aligned}
CR(f) = |\mu_{S} - \mu_{B}|,
\end{aligned}
\end{equation}
where $\mu_{S}, \mu_{B}$ denote the mean value of the structure and background, respectively.

\subsection {Baseline Algorithms}

For comparative study, the following algorithms
are used as baselines: 1) supervised learning, 2) CycleGAN \cite{zhu2017unpaired}, and 3) StarGAN \cite{choi2018stargan}.

For supervised learning, the same generator architecture was trained using a paired data set.
We normalized the intensity of the entire image from $-1$ to $1$. We also used data augmentation technique like flipping, rotating and random scaling. We trained the network with the Adam optimizer with the learning rate 1e-4. The total epoch is $200$ and the batch size is 4.

For CycleGAN training, we used the same generator architecture as the proposed method.   The discriminator also has same architecture except the last layer. There is no domain classification loss in CycleGAN. We also used WGAN with gradient penalty technique \cite{gulrajani2017improved} similar to the proposed method. 
Unlike the supervised learning,
there are no paired dataset for CycleGAN training.
We used $\lambda_{GP}$ as 30 and $\lambda_{cyc}$ as 0.5. We normalized the intensity of the image as $-1$ to $1$. To increase the total dataset, we used data augmentation technique like flipping, rotating and random scaling. We trained the network with the Adam optimizer with learning rate decreasing linearly from 1e-3 to 1e-4 depending on the training epoch. The total epoch is 200 and batch size is 4.

The StarGAN also used same generator and discriminator architecutres as our method except kernel size. It also used the WGAN with gradient penalty technique \cite{gulrajani2017improved}. Both StarGAN and the proposed method use mask vectors to separate the target domain. Therefore, each target domain image can be generated by simply changing the mask vector.  However, in contrast to our method, StarGAN also utilizes the same generator for the returning path $F_\phi$ with a mask vector.  This increases the difficulty of training.
In the StarGAN, we used $\lambda_{GP}$ as 30 and $\lambda_{cls}$ as 1 and $\lambda_{cyc} $ as 10. 
We normzliaed all the images as -1 to 1. Flipping, rotating, and random scaling also used for data augmentation technique. We trained it with Adam optimizer with 1e-4 linearly decreasing from the half of total epoch. The total epoch is 500 and batch size is 4.

\section{Experimental Results}
\label{sec:results}
To verify the performance of the proposed method, we validate the algorithm with qualitative and quantitative manners. 
%
%
\subsection {Qualitative Results}
Fig.~\ref{fig:algorithm_comparison} shows the comparison results using various algorithms. We show the in vivo images from the left thryorid and right thyroid. As shown in the figure, all the deconvolution output images show the improved contrast. Specifically, in the magnified images, it is easy to recognize that the deconvolution images have sharper structure. While the StarGAN output has slightly improved compared to the input image, the output from the proposed method is closer to the target image.  Moreover, the side to side comparison with supervised and CycleGAN approach show that the proposed
method provides qualitatively comparable results.  However, it is remarkable that the deconvolution and despeckled images using supervised learning and CycleGAN are generated from independent network trained separately, whereas the proposed method generated both images with single generator.

\begin{figure}[h]
  \center
	\includegraphics[width=8cm]{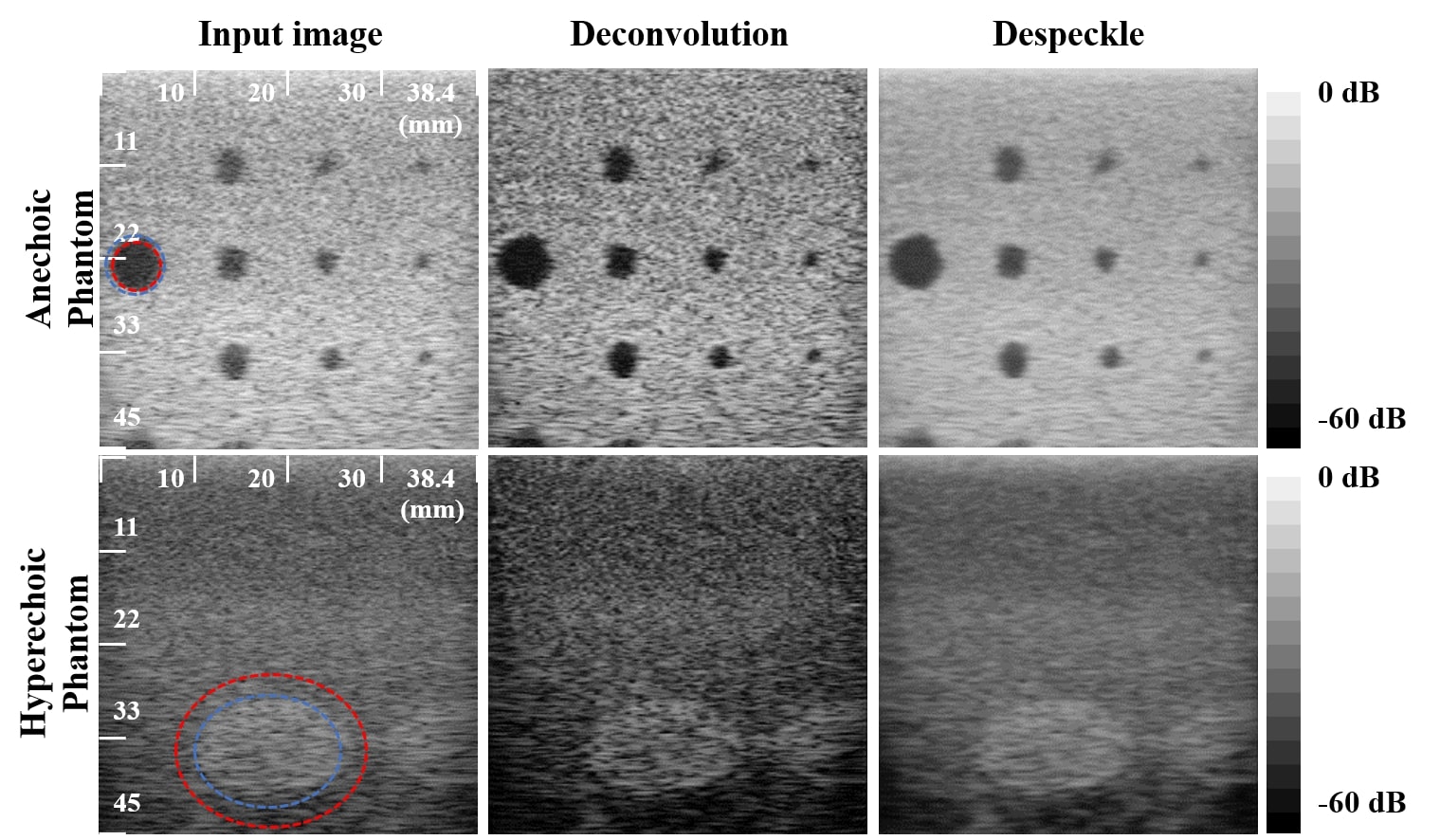}
	\vspace*{-0.3cm}
	\caption{Artifact removal in anechoic phantom and hyper-echoic phantom using our method. The red and blue line are the selected region for calculating contrast metrics. }
	\label{fig:phantom_result}
\end{figure}

The despeckle images in Fig.~\ref{fig:algorithm_comparison} also shows the noise suppression. 
 After removing the speckle noise, the boundaries of structure becomes smooth out. 
 However, as  shown in the Fig. \ref{fig:algorithm_comparison}, the result of the proposed method preserved the structure boundaries better
 than StarGAN. We applied our method to tissue mimicking phantoms. One is anechoic and the other is hyperechoic phantom. The results are shown in Fig. \ref{fig:phantom_result}. The deconvolved image in both cases relatively show high contrast compared to the input image. Moreover, the granular pattern is reduced and the boundaries of structure are well maintained in the proposed method.

\begin{figure*}[h!]
  \center
	\includegraphics[width=0.7\textwidth]{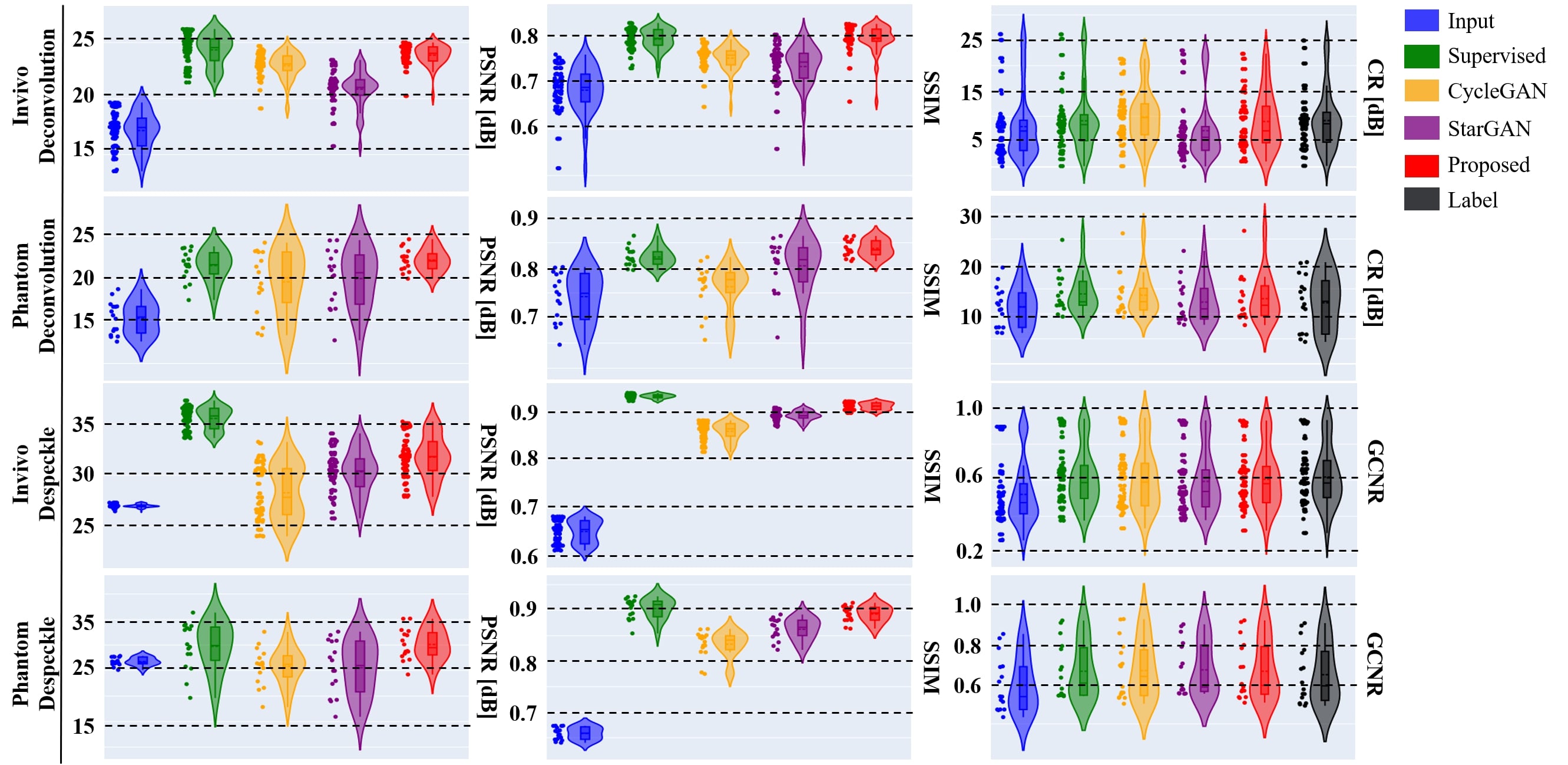}
		\vspace*{-0.3cm}
	\caption{Quantitative performance comparison. The CR graphs are shown in deconvolution case, whereas the GCNR graphs are represented in despeckle case.}
	\label{fig:performance evaluation}
\end{figure*}

\begin{table}
	\centering
		\caption{Deconvolution performance evaluation}
					\resizebox{0.5\textwidth}{!}{
	\begin{tabular}{cccccccccccc}
		\hline
		\multicolumn{2}{c}{\multirow{2}{*}{Deconvolution}} &
		\multicolumn{5}{c}{in vivo}  &
		\multicolumn{5}{c}{Phantom}   \\
		\cline{3-7} \cline{8-12}
		&    & PSNR & SSIM & CNR & GCNR & CR 
		& PSNR & SSIM & CNR & GCNR & CR \\
		\hline \hline
		\multirow{5}{*}{Algorithms}    
		&Input     	 			& 17.15     &0.68  & 0.82     &0.52  & 7.97 
		& 15.33     & 0.74  & 1.25     & 0.60 & 11.41   \\
		& Label   			  & X     & X  & 0.74     & 0.53 & 9.15
		& X    & X  & 1.13     & 0.56 & 12.24    \\
		&Supervised       & 24.35     & 0.79 & 0.79     & 0.55 & 9.13
		& 21.55    & 0.82  & 1.26     & 0.60 & 14.01 \\
		&CycleGAN       & 22.93    & 0.75  & 0.85     & 0.56 & 9.87
		& 19.62     & 0.76  & 1.19     & 0.59 & 13.75    \\
		&StarGAN       & 20.85   & 0.74  & 0.73    &  0.50 & 7.18
		&  20.02   & 0.80 &  1.17   & 0.57 &  12.38\\
		& Proposed 	  &	23.86    &  0.79   &  0.76   & 0.53 & 8.98 
		&  22.20 & 0.84 & 1.19  & 0.58  &   13.03 \\
		\hline
	\end{tabular}
	}
	\label{table:deconv_eval}
\end{table}
\begin{table}
	\centering
		\caption{Despeckle performance evaluation}
					\resizebox{0.5\textwidth}{!}{
	\begin{tabular}{cccccccccccc}
		\hline
		\multicolumn{2}{c}{\multirow{2}{*}{Despeckle}} &
		\multicolumn{5}{c}{in vivo}  &
		\multicolumn{5}{c}{Phantom}   \\
		\cline{3-7} \cline{8-12}
		&    & PSNR & SSIM & CNR & GCNR & CR 
		& PSNR & SSIM & CNR & GCNR & CR \\
		\hline \hline
		\multirow{5}{*}{Algorithms}    
		&Input     	 		  & 26.99     & 0.65  & 0.82     & 0.52 & 7.97 
		& 26.64    & 0.66  & 1.25     & 0.60 & 11.41   \\
		& Label   			  &X    	   &X  		 & 1.02     & 0.62 & 7.52
		&X     	      &X  	    & 1.46     & 0.65 & 10.79    \\
		&Supervised       & 35.58     & 0.93  & 1.04     & 0.62 & 7.29
		& 29.64     & 0.90  & 1.49     & 0.67 & 11.98 \\
		&CycleGAN       & 28.27     & 0.86  & 1.02     & 0.61 & 7.08
		& 25.72    & 0.83  & 1.51     & 0.67 & 13.83    \\
		&StarGAN       & 30.34     & 0.89 &  0.98  & 0.59 & 7.29
		&  25.57   & 0.86 &  1.47   & 0.67 &  12.55\\
		& Proposed 		   &  31.78   &  0.91 &  1.02   & 0.61 & 7.86
		&  30.11    & 0.89  &  1.49    & 0.67 &   12.52\\
		\hline
	\end{tabular}
	}
	\label{table:despeck_eval}
\end{table}

\subsection {Quantitative Results}
\label{subsec:Qunatitative}

Quantitative comparison with other methods was performed using the PSNR, SSIM, CNR, GCNR, and CR values. All test data are B-mode focused image acquired from linear probe with 8.5MHz center frequency. The test dataset are composed of in vivo and phantom images. The in vivo datasets consist of 80 frames from four parts of two subjects. The phantom datasets are totally 16 images, consisting of 10 anechoic and 6 hyperechoic phantom images. 

Fig. \ref{fig:performance evaluation} shows the distribution of each quantitative metrics. The PSNR and SSIM values show significant improvement in all algorithms. As expected, the PSNR and SSIM values of the supervised learning was the best,
but it is remarkable that the proposed method are superior to CycleGAN and StarGAN results in all cases.

The average values of all images are represented in the Table \ref{table:deconv_eval} and Table \ref{table:despeck_eval}. When the deconvolution is the target domain, in our method the PSNR value has 6.71 dB and 6.87 dB  improvement compared to the original DAS images in in vivo and phantom case, respectively. The SSIM value also gained 0.11 and 0.10 unit, respectively. Similarly, when the target domain is despeckle images, the PSNR and SSIM values show  high improvement in both in vivo and phantom images.

In the Table \ref{table:deconv_eval}, the CNR and GCNR values of our method for deconvolution tasks are slightly decreased compared to DAS input.  This is because the deconvolution process also enhance the noise component. On the other hand, the CR values, which is less sensitive to noise boosting,
 is increased. The result from the proposed method shows the CR value enhancement about 1.01dB and 1.62dB   in in vivo and phantom images, respectively. 

On the other hand, for the despeckle task, the CNR, GCNR value from our method is increased significantly. Table \ref{table:despeck_eval} shows that the results using our method has 0.2 and 0.24 unit improved CNR value for in vivo and phantom images, respectively. Similarly, the GCNR value is increased about 0.09 and 0.07 unit.

%

\subsection {Computational Time}

The average computational time for classical deconvolution model \cite{duan2016increasing}, NLLR despeckle  \cite{zhu2017non},
and our method are 3479.3sec, 435.51sec, and 0.16sec, respectively.
The computational times for supervised learning, CycleGAN, StarGAN are basically same as ours.

In conventional deconvolution algorithm \cite{duan2016increasing}, the point spread function (PSF) should be calculated to
 produce the deconvolution image. The estimation of the PSF and
 the joint sparse recovery step is computationally expensive due to iterative method.
  Furthermore, the  Non-Local-Low-Rank (NLLR) approach for classical despeckle approach is also based on the iterative method, so it requires huge
  computational burden to reconstruct the image \cite{zhu2017non}. 
  On the other hand, our method is based on deep neural network so that once the network is trained, the artifact-free images can be generated  in a real time manner.
Especially, the proposed method can generate both deconvolution image and despekcle image at the same time. 


\section{Discussion}
\label{sec:discussion}
\subsection {Generalization}
To verify the generalization power of the proposed algorithm, we tested our method with various types of dataset. The new datasets are acquired with different manner compared to the previous dataset.

\subsubsection{Planewave mode}
Although we trained the network with B-mode images acquired with focused mode, we tested our network with planewave mode images to show that the proposed method is independent of the acquisition mode.
The planewave image is acquired with linear array probe (L3-12H) and 8.5MHz center frequency. The 31 planewaves are used to generate the final images.

\begin{figure}[h!]
  \center
	\includegraphics[width=8cm]{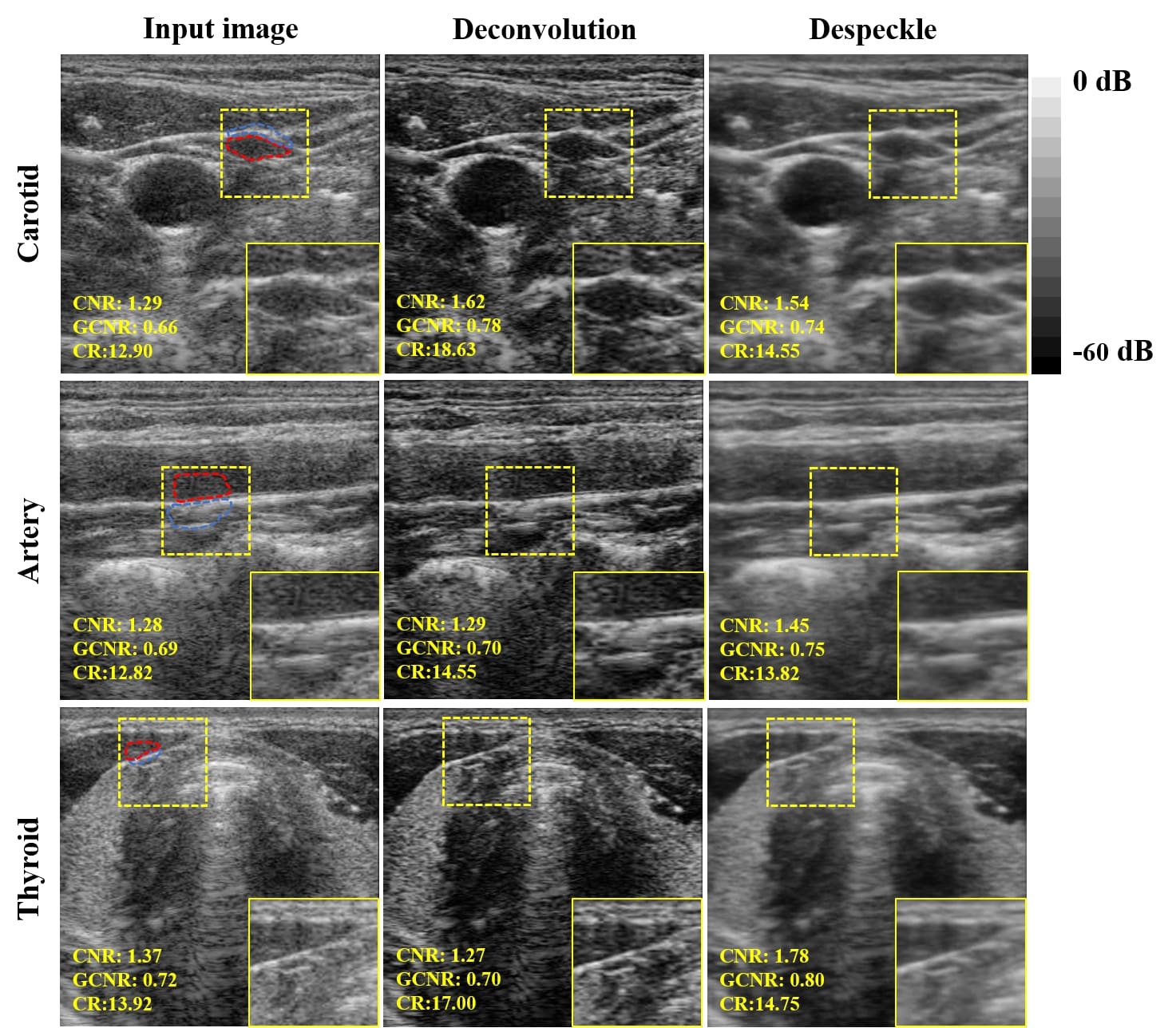}
		\vspace*{-0.3cm}
	\caption{Results of planewave mode images using our method. }
	\label{fig:planewave}
\end{figure}

As shown in Fig. \ref{fig:planewave}, the proposed network successfully deconvolve and despeckle the input images. The deconvolved images have better contrast compared to input DAS image. Our method also well suppressed the speckle noises from the planewave mode images.
The improvement is prominent in quantitative comparison. When the target domain is the deconvolution image, each image has 5.73 dB, 1.73 dB, and 3.08dB gain in the CR value. Similary, the despeckled images show 0.08 unit gain in GCNR value.

\subsubsection{Different operating frequency}
The training datasets are composed of B-mode focused images with operating freqeuncy of 8.5MHz. In this experiment, we show the generalization power with operating frequency of 10.0MHz. Moreover, the datasets are from totally different body part: forearm muscle and calf muscle.

\begin{figure}[h!]
  \center
  		\vspace*{-0.3cm}
	\includegraphics[width=8cm]{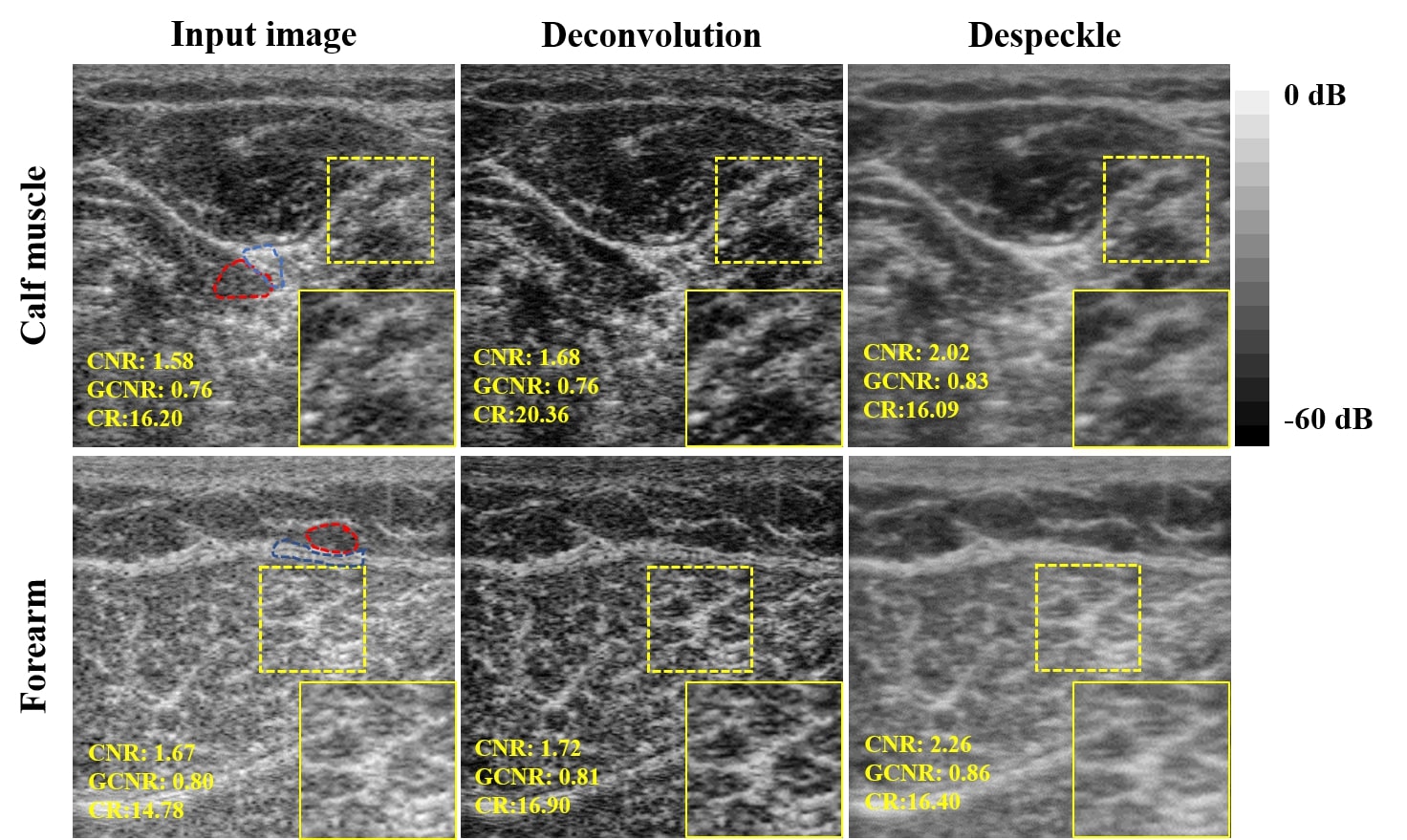}
		\vspace*{-0.3cm}
	\caption{Results of the data with the center frequency of 10.0MHz using our method. The first row is calf muscle and the second row is forearm muscle.}
	\label{fig:operaint_freq}
\end{figure}

Fig. \ref{fig:operaint_freq} shows the results. It can be easily seen that the result from the proposed method removed the artifact well. The details of structure become prominent in deconvolved image. The granular noise reduced well with maintaining the structure boundaries. 

\subsection {Extension beyond two domains}

In the experiment above, we tried to remove two types of artifact. In this section, we extended artifacts to deal with. 
That is, we increased the number of target doamain from two to three. The first target domain is deconvolution, the second target domain is despeckle and the third target domain is decovolution-despeckle. To check the effect of domain extension, we carried out the following two experiments. 
Note that except mask vector, other processes are all same as main experiment.

\begin{enumerate}
	\item Two labels for three target domains : The mask vector [1,0] is for deconvolution, [0,1] is for despeckle, [1,1] is for deconvolution-despeckle.
	\item Three labels for three target domains: The mask vector [1,0,0] is for deconvolution, [0,1,0] is for despeckle, [0,0,1] is for deconvolution-despeckle.
\end{enumerate}

\begin{figure}[h!]
  \center
  		\vspace*{-0.3cm}
	\includegraphics[width=8cm]{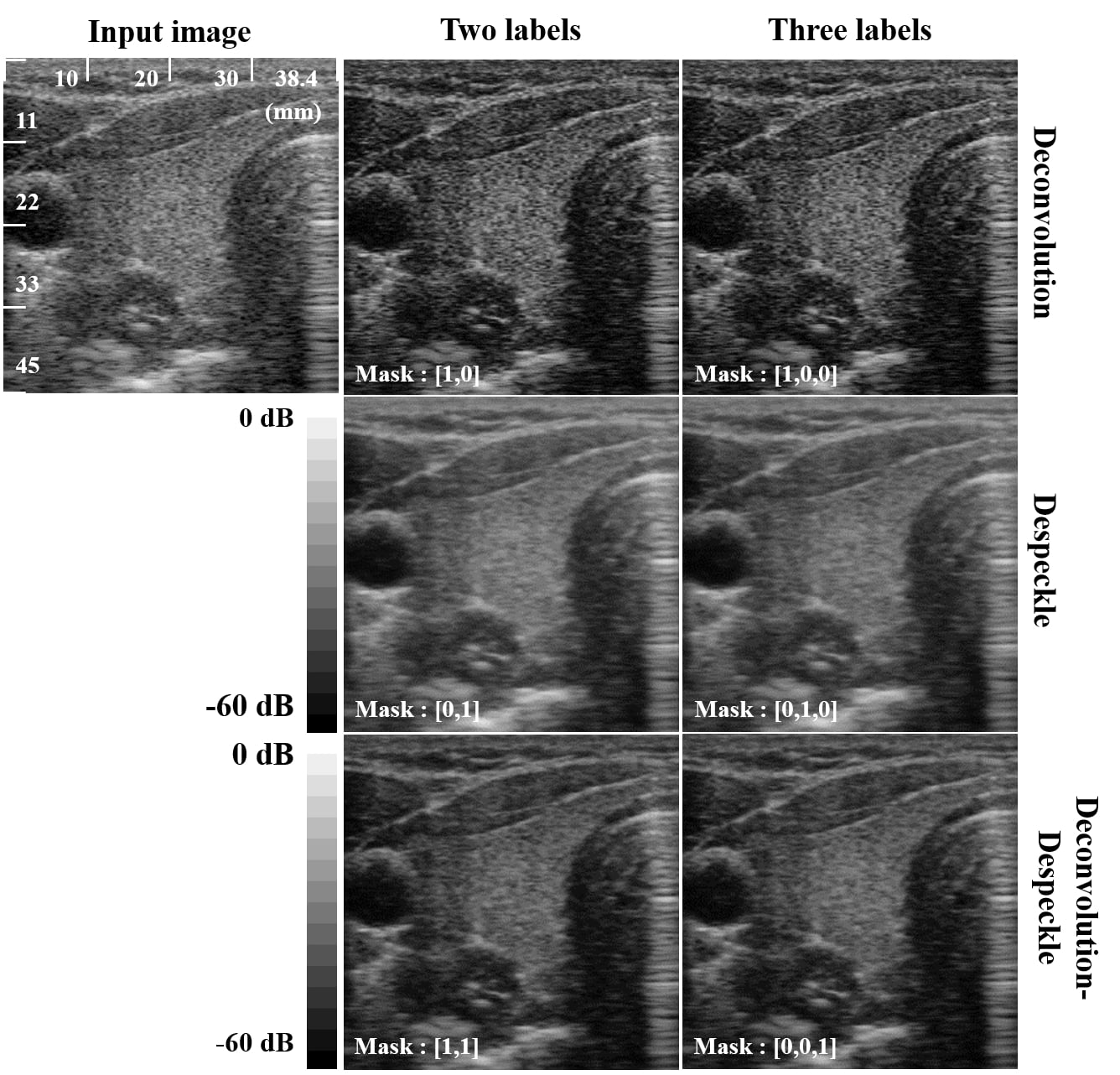}
		\vspace*{-0.3cm}
	\caption{Extension  to three target domains with different number of mask vector channels. }
	\label{fig:domain_extension}
\end{figure}

As shown in Fig. \ref{fig:domain_extension}, our method works well when increasing the number of target domain. Both deconvolved images have better contrast and the speckle noise is well suppressed in despeckle images. Furthermore, the deconvolution-despeckle images show reasonable result. Two type of masks described above generates three distinct domain similarly. Therefore, we can conclude that
extension of multiple domain beyond two can be easily done without concerning about the increase of the mask vector channels.


\section{Conclusion}
\label{sec:conclusion}
Due to the transducer limitations or wave interference, the ultrasound image has many artifacts. To overcome this problem, we proposed the multi-domain image artifact removal method based on optimal transport theory. 
Using extensive experiments using in vivo and phantom dataset,
we confirmed that a single generator can produce the multiple domain targets very well in both qualitatively and quantitatively.
As the proposed method can process images instantaneously without increasing the number of models,
 we believe that our method can be used for practical US systems.


\bibliographystyle{IEEEtran}
\bibliography{ref}

\end{document}